\documentclass{ws-mpla}

\begin{document}

\markboth{A.V. Gladyshev, R.S. Parpalak}
{$R$-Parity Violation and Supersymmetric Higgs Masses}

\catchline{}{}{}{}{}

\title{$R$-PARITY VIOLATION\\ AND SUPERSYMMETRIC HIGGS MASSES}

\author{\footnotesize A.V. GLADYSHEV, R.S. PARPALAK}

\address{Bogoliubov Laboratory of Theoretical Physics, Joint Institute for Nuclear Research,\\
6 Joliot-Curie, Dubna, Moscow Region, 141980 Russian Federation\\
gladysh@theor.jinr.ru, parpalak@dgap.mipt.ru}

\maketitle

\pub{Received (Day Month Year)}{Revised (Day Month Year)}

\begin{abstract}
We consider the supersymmetric extension of the Standard Model with neutrino Yukawa
interactions and $R$-parity violation. We found that $R$-parity breaking term 
$\lambda_i \bar\nu_i H_{\rm u} H_{\rm d}$ leads to an additional $F$-type contribution to the
Higgs scalar potential, and thus to the masses of supersymmetric Higgs bosons. The most interesting 
consequence is the modification of the tree-level expression for the lightest neutral 
supersymmetric Higgs boson mass. It appears that due to this contribution the bound 
on the lightest Higgs mass may be shifted upwards, thus slightly opening the part of the 
model parameter space excluded by non-observation of the light Higgs boson at LEP in the framework of the Minimal Supersymmetric Standard Model.
\end{abstract}

\ccode{PACS Nos.: 12.60.Jv, 14.80.Ly}

\section{Introduction}	

The searches for the Higgs boson and new phenomena beyond the Standard Model of fundumental interactions are important tasks for the new particle accelerator -- the Large Hadron Collider. The most popular direction beyond the SM is low energy supersymmetry. However, it is not clear how supersymmetry is realized. The simplest case -- the Minimal supersymmetric Standard Model (MSSM)\cite{MSSM1,MSSM2,MSSM3,MSSM4} -- is studied in detailes, however possible deviations from it are of great interest as well. There exists a wide class of models which contain so called $R$-parity breaking interactions leading to the violation of lepton and baryon numbers\cite{RPV}. These models have a number of new coupling constants, some of them are badly constrained, for instance, by rare processes, other ones are less restricted.

In this Letter we consider the model with the neutrino Yukawa interactions and related $R$-parity violating term in the superpotential. The interesting consequence of including the $R$-parity violating $\lambda_i \bar\nu_i H_{\rm u} H_{\rm d}$ term is the modification of the tree-level expressions for the Higgs boson masses and upper bound on the lightest Higgs boson mass in the MSSM.

\section{MSSM Higgs scalar potential, minimization conditions and supersymmetric Higgs boson masses}

In this chapter we consider general features of supersymmetric theories related to the scalar potential, its minima, and masses of the physical Higgs boson states. It is well known that in these theories the scalar potential is not an arbitrary function of the fields but is fixed by supersymmetry. The sources of the Higgs scalar potential are $D$-terms coming from gauge-Higgs interactions, $F$-terms originating from the superpotential, and soft supersymmetry breaking terms:
$$
V_{Higgs}= \frac12 D^2 + F^*F + V_{soft}(H)
$$
In the MSSM based on the $SU(2)_L\times U(1)_Y$ symmetry group the $D$-term contribution reads
$$
\frac{1}{2}\sum_{{\rm G}}\sum_\alpha \sum_{i, j} g_{\rm G}^2
(\phi_i^\dagger  T_{\rm G}^\alpha \phi_i) (\phi_j^\dagger T_{\rm G}^\alpha \phi_j)
$$
where the sum is taken over the gauge groups, their generators, and scalar components of the chiral Higgs superfields.
From the $R$-parity conserving MSSM superpotential
\begin{equation}
W_{RMSSM} = y_{\rm u}^{ij}{\bar{ u}}_i Q_j\! \cdot\! H_{\rm u}
    - y_{\rm d}^{ij}{\bar{ d}}_i Q_j\! \cdot\! H_{\rm d}
    - y_{\rm e}^{ij}{\bar{ e}}_i L_j\! \cdot\! H_{\rm d}
    + \mu H_{\rm u}\! \cdot\! H_{\rm d},
     \label{eq:WMSSM}
\end{equation}
only the Higgs mixing $\mu$-term
$$
W_{RMSSM} \propto \mu H_{\rm u}\! \cdot\! H_{\rm d} = \mu \left( H^+_{\rm u} H^-_{\rm d} - H^0_{\rm u} H^0_{\rm d} \right)
$$
gives the contribution to the $F$-part of the $V_{Higgs}$:
$$
\left|\frac{\partial W}{\partial H_i} \right|^2
\propto |\mu|^2 (|H^+_{\rm u}|^2 +|H^-_{\rm d}|^2 + |H^0_{\rm u}|^2 + |H^0_{\rm d}|^2).
$$
The Higgs doublets are defined as
$$
H_{\rm u} = \left( \begin{array}{c} H^+_{\rm u} \\ H^0_{\rm u} \end{array} \right), \ \ \
H_{\rm d} = \left( \begin{array}{c} H^0_{\rm d} \\ H^-_{\rm d} \end{array} \right),
$$
and the "dot" operation ($\cdot$) denotes the $SU(2)$ convolution of the (super)fields doublets
with the help of the totally antisymmetric $\epsilon_{ij}$-tensor.

Soft supersymmetry breaking is parametrized by the mass terms for the corresponding scalar components of the Higgs superfields and an analogue of the Higgs mixing term $B\mu H_{\rm d}\!\cdot\! H_{\rm u}$
\begin{equation}
V_{soft} =
-m^2_{{\rm H}_{\rm u}} H_{\rm u}^\dagger \!\cdot\! H_{\rm u} - m^2_{{\rm H}_{\rm d}}
H^\dagger_{\rm d} \!\cdot\! H_{\rm d} - (B\mu H_{\rm u} \!\cdot\! H_{\rm d} + {\rm h.c.}).
\label{eq:softhiggsmass}
\end{equation}

The final form of the Higgs scalar potential then reads
\begin{eqnarray}
V(H_{\rm d},H_{\rm u})
&=& \left(|\mu|^2 + m^2_{{\rm H}_{\rm u}}\right)\left(|H^+_{\rm u}|^2+|H^0_{\rm u}|^2\right) +
    \left(|\mu|^2 +m^2_{{\rm H}_{\rm d}}\right)\left(|H^0_{\rm d}|^2+|H^-_{\rm d}|^2\right) \nonumber \\[1mm]
&+& B\mu \left(H^+_{\rm u}H^-_{\rm d}-H^0_{\rm u}H^0_{\rm d}\right) + {\rm h.c.} \nonumber \\
&+& \frac{g^2+g^{\prime 2}}{8}\Bigl(|H^+_{\rm u}|^2 +|H^0_{\rm u}|^2-|H^0_{\rm d}|^2-|H^-_{\rm d}|^2\Bigr)^2 \nonumber \\
&+& \frac{g^2}{2}\left|H^+_{\rm u}H^{0\dagger}_{\rm d}+H^0_{\rm u}H^{-\dagger}_{\rm d}\right|^2. 
\label{eq:Vscal_MSSM}
\end{eqnarray}
To calculate the masses of physical Higgs boson states one has to take the second derivatives of the scalar potential (\ref{eq:Vscal_MSSM}) with respect to the corresponding real and imaginary components of the Higgs fields taken at the minima and then diagonalize the mass-squared matrices. Then one gets the following expressions\cite{ERZ1,ERZ2}
for the mass of the $CP$-odd Higgs boson $A^0$
\begin{equation}
m_{A^0}^2 = \frac{2B\mu}{\sin 2\beta} ,
\end{equation}
for the heavy and the lightest neutral $CP$-even Higgs bosons $H^0$ and $h^0$
\begin{equation}
m_{H^0,h^0}^2 = \frac12 \left( m^2_{A^0} +M^2_Z \pm \sqrt{\left( m^2_{A^0} + M^2_Z \right)^2
- 4 m^2_{A^0} M^2_Z \cos^2 2 \beta} \, \right),
\label{eq:massHh}
\end{equation}
and for the pair of charged Higgs bosons $H^\pm$
\begin{equation}
m_{H^\pm}^2 = m_{A^0}^2 + M_W^2.
\end{equation}
Three mass eigenstates remain zero corresponding to the Goldstone bosons eaten by $SU(2)_L$ gauge bosons in the Higgs mechanism. Here we have introduced the standard notation $\tan\beta=v_{\rm u}/v_{\rm d}$ --- the ratio of the vacuum expectation values of the neutral components of the two Higgs boson doublets.

The assumption $m_{A^0}^2 \gg M_{Z^0}^2$ in Eqn.(\ref{eq:massHh}) leads to the well-known tree-level expression for the lightest MSSM Higgs boson mass
\begin{equation}
m_{h^0}^2 = M_{Z^0}^2 \cos^2 2\beta
\end{equation}
and the famous inequality
\begin{equation}
m_{h^0}^2  < M_{Z^0}^2.
\end{equation}
However, radiative corrections coming mainly from the top-stop loops badly violate this inequality\cite{ERZ1}.
For the loop corrected lightest Higgs mass one has
\begin{equation}
m_{h^0}^2 = M_{Z^0}^2 \cos^2 2\beta + \frac{3g^2m_t^4}{16\pi^2M_W^2}
\log\frac{\tilde m^2_{t_1}\tilde m^2_{t_2}}{m_t^4}p + {\rm 2\ loops}
\end{equation}
which shift the mass bound upwards\cite{GKBBE,PBMZ,CQW,2-loops_higgs_1,2-loops_higgs_2}. Renormgroup resummation of all-order leading log contributions using effective potential approach slightly changes the predictions\cite{GK}.

Besides, the MSSM scenario with $\tan\beta \leq 3$ is excluded experimentally by non-observation of the Higgs boson lighter than 114~GeV.

\section{Supersymmetric Standard Model with the Right-handed Neutrino and $R$-parity breaking}

The superpotential of the Standard Model~(\ref{eq:WMSSM}) is constructed under assumption that neutrinos are massless (there are no Yukawa interactions for the neutrinos which can generate Dirac neutrino mass terms after the electroweak symmetry breaking) and the $R$-parity is conserved. In this case it repeats (up to notations) the Yukawa part of the Standard Model lagrangian. However, it is believed nowadays that neutrinos have masses, even tiny, then the neutrino Yukawa term
\begin{equation}
y_{\rm \nu}^{ij} {\bar{\nu}}_i  L_j \!\cdot\! H_{\rm u}
\label{eq:W_nu_r}
\end{equation}
in the superpotential is possible and should be included ($\bar\nu_i$ here are $SU(2)$ singlet right-handed neutrino superfields, and $y_{\rm \nu}^{ij}$ are neutrino Yukawa couplings). The latter implies that one also has to include a term
\begin{equation}
\lambda_\nu^i \bar\nu_i H_{\rm u} \!\cdot\! H_{\rm d}.
\label{eq:W_nu_nr}
\end{equation}
to the $R$-violating part (recall that $L$ and $H_{\rm u}$ superfields have the same quantum numbers and no symmetries but lepton number are violated). Therefore we consider the superpotential of the model in the form:
\begin{equation}
W = W_{RMSSM}
+ y_\nu^{ij} \bar\nu_i  L_j \!\cdot\! H_{\rm u}
+ \lambda_\nu^i \bar\nu_i H_{\rm u} \!\cdot\! H_{\rm d}.
\end{equation}

The soft supersymmetry breaking Lagrangian also includes the following terms:
\begin{equation}
{\cal L}_{SSB} = ...
+ A_\nu^{ij} \bar\nu_i  L_j \!\cdot\! H_{\rm u}
+ A_\lambda^i \bar\nu_i H_{\rm u} \!\cdot\! H_{\rm d}.
\end{equation}
($\bar\nu_i,  L_j, H_{\rm u}, H_{\rm d}$ here are scalar components of the corresponding superfields).

The model with this kind of superpotential has been previously considered and studied\cite{munoz,roy}, however the authors were mainly interested in the solution of the $\mu$-problem rather than Higgs mass predictions in the model.
Note also, that for simplicity we do not consider sneutrino v.e.v.

The $\lambda_\nu^i \bar\nu_i H_{\rm u} \!\cdot\! H_{\rm d}$ term gives the $F$-type contribution to the Higgs self-coupling, and the Higgs scalar potential now reads
\begin{eqnarray}
V &=& V_{MSSM} + \left|\lambda_\nu^i \lambda_\nu^i \right| \left| H^+_{\rm u} H^-_{\rm d} - H^0_{\rm u} H^0_{\rm d} \right|^2 \nonumber \\[1mm]
&=& \left(|\mu|^2 + m^2_{{\rm H}_{\rm u}}\right)\left(|H^+_{\rm u}|^2+|H^0_{\rm u}|^2\right) +
    \left(|\mu|^2 +m^2_{{\rm H}_{\rm d}}\right)\left(|H^0_{\rm d}|^2+|H^-_{\rm d}|^2\right) \nonumber \\[1mm]
&+& B\mu \left(H^+_{\rm u}H^-_{\rm d}-H^0_{\rm u}H^0_{\rm d}\right) + {\rm h.c.} \nonumber \\
&+& \frac{g^2+g^{\prime 2}}{8}\Bigl(|H^+_{\rm u}|^2 +|H^0_{\rm u}|^2-|H^0_{\rm d}|^2-|H^-_{\rm d}|^2\Bigr)^2
    + \frac{g^2}{2}\left|H^+_{\rm u}H^{0\dagger}_{\rm d}+H^0_{\rm u}H^{-\dagger}_{\rm d}\right|^2 \nonumber \\
&+& \left|\lambda_\nu^i \lambda_\nu^i \right| \left| H^+_{\rm u} H^-_{\rm d} - H^0_{\rm u} H^0_{\rm d} \right|^2. 
\label{eq:VscalHR}
\end{eqnarray}
The minimization conditions for the potential (\ref{eq:VscalHR}) have an additional term due to new neutrino-Higgs interaction (the notation $\varepsilon^2 = \left| \lambda_\nu^i \lambda_\nu^i \right|/(g^2+g^{\prime 2})$ is introduced):
\begin{eqnarray}
|\mu|^2 + m^2_{{\rm H}_{\rm u}} &=& B\mu \cot \beta + \frac{m^2_{Z^0}}{2} \cos 2 \beta - 2 \varepsilon^2 M^2_{Z^0} \cos^2 \beta
\label{eq:mincon1} \\
|\mu|^2 + m^2_{{\rm H}_{\rm d}} &=& B\mu \tan \beta - \frac{m^2_{Z^0}}{2} \cos 2 \beta - 2 \varepsilon^2 M^2_{Z^0} \sin^2 \beta.
\label{eq:mincon2}
\end{eqnarray}

\section{Higgs Boson Masses}

Now it is easy to calculate the masses of the Higgs bosons in our model.

Taking the second derivatives of the potential~(\ref{eq:VscalHR}) with respect to ${\rm Im} H^0_{\rm u}$ and ${\rm Im} H^0_{\rm u}$ in the minimum one gets for the elements of the $CP$-odd Higgs boson mass-squared matrix:
\begin{eqnarray}
M_{11}^{\rm sq} &=& |\mu|^2 +m^2_{{\rm H}_{\rm u}}+
\frac{g^2+g^{\prime 2}}{4}(v^2_{\rm u}-v^2_{\rm d}) + \lambda^2v^2_{\rm d} = B\mu \cot \beta, \nonumber \\
M_{12}^{\rm sq} &=& M_{21}^{\rm sq} = B\mu, \nonumber \\
M_{22}^{\rm sq} &=& B\mu \tan \beta.
\label{eq:A0matrix11}
\end{eqnarray}
The mass eigenstates are then given by
$$
m^2_1= \frac{2B\mu}{\sin 2\beta}, \quad m^2_2=0.
$$
The zero eigenstate of the matrix (\ref{eq:A0matrix11})
$$
\sqrt{2}[\sin \beta ({\rm Im}H^0_{\rm u}) - \cos \beta ({\rm Im}H^0_{\rm d})].
$$
is the the Goldstone boson eaten by the $Z^0$ boson in the Higgs mechanism, while the combination
$$
\sqrt{2}[\cos \beta ({\rm Im} H^0_{\rm u}) + \sin \beta ({\rm Im} H^0_{\rm d})]
$$
is the $CP$-odd neutral Higgs boson $A^0$ with the mass expression
\begin{equation}
m^2_{A^0} = \frac{2B\mu}{\sin 2\beta}.
\label{eq:massA0}
\end{equation}
Sometimes it is more convenient to use $m_{A^0}$ instead of the $B$ parameter, they are related via (\ref{eq:massA0}).

Let us now look at the masses of $CP$-even Higgs boson $H$ and $h$, the latter corresponds to the lightest Higgs boson in the MSSM. The mass-squared matrix is obtained by differentiating twice the potential~(\ref{eq:VscalHR}) with respect to the ${\rm Re} H^0_{\rm u}$ and ${\rm Re} H^0_{\rm d}$ and substituting the expression for the vacuum expectation values (\ref{eq:mincon1}) and (\ref{eq:mincon2}). One has, for instance
\begin{eqnarray}
M_{11}^{\rm sq} &=& |\mu|^2 +m^2_{{\rm H}_{\rm u}}+
\frac{g^2+g^{\prime 2}}{4}(2v^2_{\rm u}-v^2_{\rm d}) + \lambda^2v^2_{\rm d} = B\mu \cot\beta + M^2_{Z^0} \sin^2 \beta, \\
M_{12}^{\rm sq} &=& -B\mu -
\frac{g^2+g^{\prime 2}}{2}v_{\rm u}v_{\rm d} + 2 \lambda^2v_{\rm u}v_{\rm d} = -B\mu - \frac{1}{2} M^2_{Z^0} ( 1 - 4 \varepsilon^2 )  \sin 2\beta .
\end{eqnarray}
The neutral $CP$-even Higgs boson mass-squared matrix then reads
\begin{equation}
{\bf M}^{\rm sq}_{{\rm H, h}}=
\left( \begin{array}{cc}
B\mu \cot\beta + M^2_{Z^0} \sin^2 \beta & -B\mu - \frac{1}{2} M^2_{Z^0} \left( 1 - 4 \varepsilon^2 \right) \sin 2\beta \\
-B\mu -\frac{1}{2} M^2_{Z^0} \left( 1 - 4 \varepsilon^2 \right) \sin 2\beta & B\mu \tan\beta + M^2_{Z^0} \cos^2\beta \end{array}
\right).
\label{eq:CPeven_matrix}
\end{equation}

The masses of physical eigenstates are then given by
\begin{equation}
m^2_{H,h}= \frac{1}{2} \left( m^2_{A^0} +M^2_{Z^0} \pm \sqrt{(m^2_{A^0}+M^2_{Z^0})^2
-4 m^2_{A^0} M^2_{Z^0} \cos^2 2\beta + \Delta_\varepsilon} \right),
\label{eq:massHh}
\end{equation}
where the quantity $\Delta_\varepsilon$ related to the new $R$-violating neutrino-Higgs interactions considered above reads
$$
\Delta_\varepsilon = - 8 m^2_{A^0} M^2_{Z^0} \varepsilon^2 \sin^2 2\beta -
8 M^4_{Z^0} \varepsilon^2 (1 - 2 \varepsilon^2) \sin^2 2\beta.
$$
In the limiting case $m_{A^0}\ll M_{Z^0}$ one gets
\begin{equation}
m^2_h = M^2_{Z^0} \left( \cos^2 2\beta + 2\varepsilon^2 \sin^2 2\beta \right)
\label{eq:massh_nu}
\end{equation}
which reproduces the MSSM tree-level upper bound for the lightest MSSM Higgs boson at $\varepsilon=0$
\begin{equation}
m^2_h = M^2_{Z^0} \cos^2 2\beta < M^2_{Z^0}.
\label{eq:massh_susy}
\end{equation}

\begin{figure}[bt]
\centerline{\psfig{file=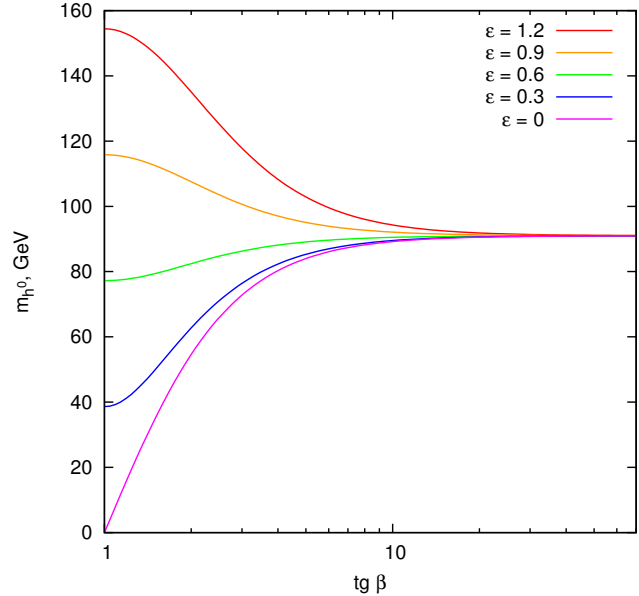,width=0.7\textwidth}}
\vspace*{8pt}
\caption{The dependence of the lightest Higgs boson mass on $\tan\beta$ for different values of $\varepsilon$.
\protect\label{fig:hmass_vs_tan}}
\end{figure}

\begin{figure}[bt]
\centerline{\psfig{file=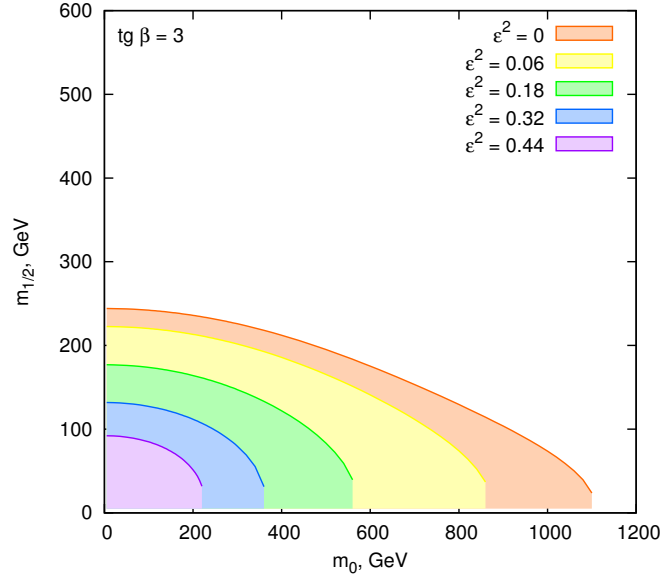,width=0.7\textwidth}}
\vspace*{8pt}
\caption{Regions of the parameter space excluded by the non-observation of the Higgs boson for different values of $\varepsilon$ .
\protect\label{fig:param_space}}
\end{figure}

Thus we conclude that in our model the Higgs mass constraint is less stringent. Fig.\ref{fig:hmass_vs_tan} illustrates the dependence of the upper bound for the lightest supersymmetric Higgs mass as a function of $\tan\beta$ for various values of the $\varepsilon$ parameter. It is easily observed that the bigger the value of $\varepsilon$, the higher the mass bound. In particular, this takes place for small values of $\tan\beta$, thus making the parameter space less constrained, and slightly open the scenario with low $\tan\beta$ excluded in the MSSM by non-observation of the Higgs boson lighter than 114~GeV~(Fig.\ref{fig:param_space}). For the moderate and large $\tan\beta > 10$ the contribution proportional to $\varepsilon^2$ is practically negligible, and the Higgs mass bound as well as the excluded region in the parameter space practically coincides with those of the MSSM. 

Note, that the modified upper bound on the lightest Higgs mass~(\ref{eq:massh_nu}) has the same form as in the Next-to-minimal supersymmetric Standard Model (NMSSM)
\begin{equation}
m_h^2 = M^2_{Z^0}\cos^2 2\beta + \kappa^2 v^2 \sin^2 2\beta,
\label{eq:masshNMSSM}
\end{equation}
where $\kappa$ is the coupling of the three $SU(2)$ singlet Higgs superfields.

For completeness we also calculate the charged Higgs boson masses which get negative contribution from the $\lambda^i_\nu \bar\nu_i H_{\rm u} H_{\rm d}$ term. The charged Higgs mass-squared matrix reads
$$
{\bf M}^{\rm sq}_{{\rm H^\pm}}=
\left(B\mu+(\frac{g^2}{2}-|\lambda_i \lambda_i|) v^2 \sin\beta \cos\beta \right)
\left( \begin{array}{cc}
\cot\beta & 1 \\
1 & \tan\beta \end{array}
\right).
$$
and the physical state mass is
\begin{equation}
m^2_{\rm H^\pm} = m^2_{A^0} + M^2_W - 2\varepsilon^2 M^2_Z,
\label{eq:massHch}
\end{equation}
The zero mass eigenstate corresponds to the Goldstone boson eaten by $W$-boson.

We have not considered in this Letter the possible bilinear $R$-parity violating term $\mu_i L_i H_{\rm u}$ in the superpotential, since for our purposes it is practically irrelevant. Its presence leads only to the redefinition of the Higgs scalar potential parameters, while the Higgs mass relations~(\ref{eq:massA0}), (\ref{eq:massHh}), and (\ref{eq:massHch}) remain unchanged. Our goal is to demonstrate the possibility of relaxing the upper bound on the lightest supersymmetric Higgs.

\section{Conclusions}

In Conclusion we summarize the main results of this Letter. We have shown that introducing the Yukawa interactions of neutrinos $y_{\rm \nu}^{ij} {\bar{\nu}}_i  L_j H_{\rm u}$ leading to their Dirac mass terms and consequently the possible $R$-parity violating terms $\lambda_\nu^i {\bar{\nu}}_i H_{\rm u} H_{\rm d}$ modifies the expressions of the Higgs boson masses since the latter gives $F$-type contribution to the Higgs scalar potential. The final expressions for the Higgs masses in our toy model are (\ref{eq:massA0}), (\ref{eq:massHh}), (\ref{eq:massHch})
and the tree-level bound on the lightest supersymmetric Higgs boson 
\begin{equation}
m^2_{{\rm h}^0} = M^2_{Z^0} \left( \cos^2 2\beta + 2\varepsilon^2 \sin^2 2\beta \right)
\end{equation}
gets the noticable shift upwards for small values of $\tan\beta$. The shift is compatible with the loop corrections to the tree-level MSSM expression. This opens the part of the parameter space excluded by the non-observation of the light Higgs. The situation looks like as in the NMSSM (\ref{eq:masshNMSSM}). Including the bilinear $R$-parity violating term $\mu_i L_i H_{\rm u}$ will not crucially affect our results. However, the more detailed study is needed, and the analysis of the full $R$-parity violating model with the right-handed neutrinos is under study and will be published elsewhere.

\section*{Acknowledgments}

Authors are grateful to D.I. Kazakov and A.V. Bednyakov for fruitful discussions.
Financial support from the Russian Foundation for Basic Research (grant \# 08-02-00856) and the Ministry of Education
and Science of the Russian Federation (grant \# 3810.2010.2) is kindly acknowledged.


\end{document}